\documentclass[12pt]{iopart}
\usepackage{iopams}
\usepackage{graphicx}
\begin{document}

\title[Canalization and symmetry in Boolean models...]{Canalization and 
symmetry in Boolean models for genetic regulatory networks}

\author{C.J. Olson Reichhardt$^{1}$ and Kevin E. Bassler$^{2,3}$}

\address{$^1$Theoretical Division and Center for Nonlinear Studies,
Los Alamos National Laboratory, Los Alamos, NM 87545}
\address{$^2$Department of Physics, University of Houston, 
Houston, TX 77204-5005}
\address{$^3$Texas Center for Superconductivity at the University of 
Houston, Houston, TX 77204}
\ead{cjrx@lanl.gov}

\begin{abstract}
Canalization of genetic regulatory networks has been argued to be
favored by evolutionary processes due to the stability that it can
confer to phenotype expression.  We explore whether a significant
amount of canalization and partial canalization can arise in purely
random networks in the absence of evolutionary pressures.
We use a mapping of the Boolean functions in the Kauffman N-K
model for genetic regulatory networks
onto a $k-$dimensional Ising
hypercube (where $k=K$) 
to show that the functions can be divided into different
classes strictly due to geometrical constraints. 
The classes can be counted and their properties determined
using results from group theory and isomer chemistry.  We 
demonstrate that
partially canalizing functions completely dominate all possible Boolean
functions,
particularly for higher $k$.  This indicates that partial canalization is
extremely common, even in randomly chosen networks, and has implications
for how much information can be obtained in experiments on native state
genetic regulatory networks.
\end{abstract}
\maketitle

\section{Introduction}

To preserve the identity of a species, biological organisms must be
capable of maintaining relatively stable phenotype expression in the face
of a variety of environmental factors and a certain level of genetic
randomness.  
Experimental observations have shown that certain developmental
traits appear to control the expression of other traits.
Waddington \cite{Waddington} termed the control of one trait by another
``canalization,'' a name derived from the analogy that the developmental
pathway of the organism is like one particular canal in a floodplain,
and the further development of the organism
is completely constrained by that canal.
Canalization produces robustness because it suppresses
those changes in phenotype expression that would require development to 
deviate from the canalized pathway. For this reason it has been 
suggested that organisms evolve to be canalized.

The significance of canalization and how it might evolve remains a 
subject of debate \cite{Gibson00}.  
Since canalization suppresses the expression of
genetic variability, experimental detection of the existence of a
canalized trait
generally involves perturbing the organism out of the canalized state
\cite{Scharloo91}.  There is good evidence for the existence of
canalization in a variety of organisms \cite{Wagner97,Debat01,Meiklejohn02};
however, the microscopic mechanism for canalization is not well established.
Presumably canalization is produced genetically by the
complex interactions between genes known as the genetic regulatory network
(GRN).
As we shall see, however, a certain amount of canalization is expected 
to appear in
GRNs even
in the absence of an 
evolutionary preference for canalization. An open question 
is whether or not real GRNs contain 
more canalization than would be expected from a random graph, 
which could indicate that evolution favors canalization.

Genetic regulatory networks 
have been proposed as the mechanisms through which identical
genetic information is expressed as different cell types within the same
organism, and they can also control distinct stages in the life cycle of
an individual cell.  Depending on the conditions experienced by a given
cell and the regulatory interactions between genes, at
any moment a distinct subset of all possible genes are activated. The state or
temporal pattern of expression produces particular cell types. Organisms
with larger numbers of genes have a larger number of potentially realizable
cell types.  There has been a recent dramatic increase in the amount of
experimental information available on the structure of genetic regulatory
networks in a range of organisms, including {\it E. coli} 
\cite{Thieffry98}, budding yeast {\it S. cervisiae} \cite{Lee02,Tong04},
{\it Drosophila} species \cite{Gibson96}, {\it Xenopus} \cite{Koide05},
and the embryo of the sea urchin {\it S. purpuratus} \cite{Davidson02}.
In the simplest representation, the nodes of the network are genes and the links between genes describe
their interactions. Generally, the interactions are directional, so that the
expression of gene
A may depend on, that is ``listen to,'' the expression of gene B, but the 
expression
of gene B doesn't necessarily depend directly on the expression of gene A. 
The connectivity of a gene indicates how many other genes it 
``listens to'' when determining whether to be in an active or inactive
state.  Analysis of the connectivity of {\it E. coli} \cite{Babu03,Ma04}
and other GRNs shows a broad distribution of connectivity
among the genes, with a significant amount of negative autoregulation.
In the context of canalization, several questions arise.  What types of
structures in a GRN produce canalization of
a trait?  Do these structures arise randomly, or do they only appear
because of a special evolutionary preference?  
How significant is canalization on the scale of the
entire regulatory network?

The easiest way to approach such questions is through a simplified
model for a genetic regulatory network such as the Kauffman N-K model
\cite{Kauffman69}, which
represents the GRN as a random Boolean network.  The N-K model
has been studied extensively 
\cite{Albert00,Coppersmith,Bilke02,Kauffman03,Socolar03,Samuelsson03,Moreira05,Drossel05a}. 
Certain features of real GRNs, including
the ability of a single network to produce multiple
cell types (which appear as multiple attractors for the network), are
captured by the N-K model.
In 
this model, each gene is 
represented by a single binary element which can be either on or off
in the state 1 or 0.  Every gene receives input from 
a fixed set of $k$ other genes that are randomly chosen when the 
network is constructed, where $k=K$.
Depending on the states of its input genes, a given gene determines
whether to express the state 1 or 0 according to a randomly chosen
Boolean function of $k$ variables. An example of a Boolean function 
for $k=3$ is
given in table 1. The value of $k$ may vary from gene to
gene. The system evolves in discrete time steps and all
genes update their states simultaneously.  The entire network eventually
settles into
an attractor cycle which produces a specific 
series of network states as a function
of time.  The initial conditions of the states of the genes 
in the network determine which
of the available attractors the network will reach.  
The different attractors are interpreted as
representing different cell types expressed by a given set of genes.

\begin{table}
\caption{An example of a $k=3$ Boolean function.}
\begin{indented}
\item[]\begin{tabular}{@{}cccc}
\br
in$_1$ & in$_2$ & in$_3$ & out \\ 
\mr
0 & 0 & 0 & 0 \\
0 & 0 & 1 & 1 \\
0 & 1 & 0 & 1 \\
0 & 1 & 1 & 0 \\
1 & 0 & 0 & 0 \\
1 & 0 & 1 & 1 \\
1 & 1 & 0 & 0 \\
1 & 1 & 1 & 1 \\
\br
\end{tabular}
\end{indented}
\end{table}

A gene with connectivity $k$ employs one of the $2^{2^k}$ possible Boolean
functions to determine its response to its $k$ inputs.  Canalization occurs
in a Boolean function if the output of the gene is fixed by a particular
value of at least one of its input genes, regardless of the values of
any other inputs to that gene.
In this case the input that fixes the output of the regulated
gene is a {\em canalizing input}. Note that one value of an input 
gene, say value 0,
can be a canalizing input even if the other 
value 1 from the same input gene is not canalizing.
Canalization also occurs in a Boolean function if particular values of two 
or more
inputs together suffice to guarantee the next state of the 
regulated gene, regardless of
the values of any other inputs to the gene. 
In this case, the inputs that together
fix the output of the regulated gene 
are said to be {\em collectively canalizing inputs}.
How canalizing a particular Boolean function is can be quantified by the 
set of 
numbers $P_n$, $n=0,1,\ldots,k-1$, which are  
the fraction of sets of $n$ individual input values 
that are canalizing or collectively canalizing.
Note that Boolean functions with $P_0=1$ have a fixed output state 
regardless of
their inputs.
Boolean functions can also be characterized by their 
internal homogeneity $p$ which is defined as the  
fraction of 1s or 0s, whichever is larger, output by the function
due to all of the possible sets of input \cite{Walker}.  

A consequence of canalization is that some of the interactions between
genes may become irrelevant.
As an extreme example, if the Boolean function of a particular gene has
$P_0=1$, this gene will be
insensitive to the state of the rest of the network and its interactions
with its input genes are irrelevant.  
The number of 
canalizing functions as a function of $k$ was
derived recently in Ref.~\cite{Just04}.
Although the behaviour of 
canalizing functions would
certainly contribute to the stability of a network that is subjected to random
perturbations, such an extreme behaviour has a detrimental effect on the
ability of the network to respond to changing conditions.  
In contrast, other 
Boolean functions successfully maintain a degree of stability while retaining
the ability to change.  For these Boolean functions, 
which we will refer to as ``partially
canalizing,'' the gene may ignore one or more of its inputs under certain
conditions.  In some cases, the gene completely ignores $n$ inputs at all 
times, so that its effective connectivity is $k_{eff}=k-n$.  In other cases,
if a particular input has the value 1, for example, the gene ignores
the remaining inputs, but if that same input has the value 0, the gene
listens to its other inputs.  Here, the effective connectivity of the gene
depends on the current state of the network.  More complex categories are
also possible, such as the nested canalizing functions 
proposed by Kauffman
\cite{Kauffman03,Kauffman04}.
Since the fraction of 
canalizing functions drops rapidly with $k$, as shown in
Ref.~\cite{Raeymaekers02}, it has been assumed that canalization plays a less
important role at high $k$ compared to small $k$ \cite{Kauffman84}.  
The class of partially
canalizing functions is considerably larger than the class of 
canalizing functions; 
however, is it large enough to dominate all classes of
functions?
As we will show below on mathematical grounds, 
the partially canalizing functions completely dominate
the class of all possible Boolean functions
as $k$ increases, so that the emergence of canalization
is essentially unavoidable in a complex network.

\section{Results}

Our approach is to examine the properties of individual Boolean functions
and to determine the amount of canalization expected from a random sample of
functions. 
Since the number of possible Boolean functions explodes 
combinatorially with $k$,
we employ powerful techniques from group theory and isomer chemistry to 
classify
the various functions and help obtain their properties.  We provide results 
through $k=5$ with these methods. 
The techniques can be applied readily to higher $k$, but become 
increasingly complicated.
Therefore, to find results for larger $k$ through $k=8$
we employ statistical sampling
methods. 

For small enough $k$, the canalization properties of the functions can be 
obtained directly from inspection.  When there are two inputs for each gene,
$k=2$, as shown in table 2, 
there are only 16 possible functions which fall into four classes:
fixed (or completely canalizing) with $P_0=1$; 
sensitive to both inputs with $P_0=0$ and $P_1=0$; and the partially
canalizing cases
with $P_0=0$ and $P_1=1/2$: 
sensitive to only one input; sensitive to one or two
inputs depending on the value of one input.

\begin{table}
\caption{The sixteen $k=2$ functions and their division into four classes.}
\begin{indented}
\item[]\begin{tabular}{@{}cccccccccccccccccccc}
\br
in &\centre{2}{A} & &\centre{4}{B} & &\centre{8}{C} & &\centre{2}{D} \\ 
\mr
00 & 0 & 1 & & 0 & 1 & 0 & 1 & & 1 & 0 & 0 & 0 & 0 & 1 & 1 & 1 & & 1 & 0\\
01 & 0 & 1 & & 0 & 1 & 1 & 0 & & 0 & 1 & 0 & 0 & 1 & 0 & 1 & 1 & & 0 & 1\\
10 & 0 & 1 & & 1 & 0 & 0 & 1 & & 0 & 0 & 1 & 0 & 1 & 1 & 0 & 1 & & 0 & 1\\
11 & 0 & 1 & & 1 & 0 & 1 & 0 & & 0 & 0 & 0 & 1 & 1 & 1 & 1 & 0 & & 1 & 0\\
\mr
$P_0$ & 1 & 1 & & 0 & 0 & 0 & 0 & & 0 & 0 & 0 & 0 & 0 & 0 & 0 & 0 & & 0 & 0\\
$P_1$ & 1 & 1 & & $\frac{1}{2}$ & $\frac{1}{2}$ & $\frac{1}{2}$ 
& $\frac{1}{2}$ & & $\frac{1}{2}$ & $\frac{1}{2}$ & $\frac{1}{2}$ 
& $\frac{1}{2}$ & $\frac{1}{2}$ & $\frac{1}{2}$ & $\frac{1}{2}$ & 
$\frac{1}{2}$ & & 0 & 0\\
\br
\end{tabular}
\end{indented}
\end{table}

Inspection becomes a less viable option as $k$ increases.  In a simulation
study of the evolution properties of the different Boolean functions, 
Bassler {\it et al.} \cite{Bassler04} observed that functions with $k=3$ inputs fell into 
14 distinct classes.  In their study all of the functions 
that were members of the same class
evolved, on average, with equal probability. 
Upon examining representative functions from each
class, they categorized the functions according to their canalization
properties $P_n$.  
The triple of numbers $P_0$, $P_1$, and $P_2$ possible for $k=3$ was nearly
enough, but not quite enough, to distinctly identify each class of function.
Whether the function was symmetric about its midpoint also needed to be
considered in defining the classes.
These observations about the structures of the functions belonging to each
class were essentially empirical.
Class membership could be important for determining the
properties of real networks since we expect that on average 
all functions in the same class will evolve with equal probability.

Here, we demonstrate that there is a fundamental geometric reason for the
existence of distinct function classes.  In the N-K model, a given
function is normally represented by a Boolean string of numbers, such as
1001, of length 
$2^{2^k}$.  Comparing different functions
by inspection amounts to comparing strings of numbers with each other.
Instead of using this representation, we consider an alternative, 
equivalent representation of each function
in the form of a unit $k$-dimensional Ising hypercube.  Each axis of the 
$k$-dimensional hypercube (or simply a $k$-hypercube)
represents one
of the $k$ input variables.  The coordinates on a given axis indicate the
state of the corresponding input variable.  
Each vertex of the $k$-hypercube represents
an output state of the gene.  
In figure 1 we illustrate the mapping of the input states onto a square 
and cube for $k=2$ and $k=3$, respectively.
The output state of the gene corresponding to an input represented by
a particular vertex can be indicated by colouring the vertex white or
black to represent the values 0 or 1.
It is important to note that this system obeys parity symmetry: replacing all
0's with 1's and vice versa results in the same canalization properties
for the function.

\begin{figure}
\includegraphics[width=0.4\textwidth]{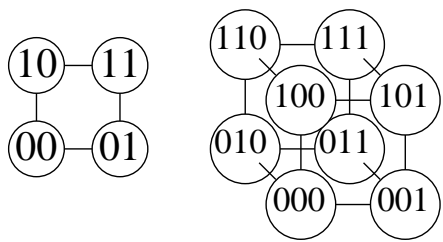}
\caption{Left: Mapping of the four possible input states for $k=2$ onto the
vertices of a square.  Right: Mapping of the eight possible input states
for $k=3$ onto the vertices of a cube.}
\end{figure}

With this hypercube mapping, it becomes clear that
functions which belong to the same class {\it have the geometric property that
they can be rotated into each other by symmetry operations on the 
$k$-hypercube plus parity.}  
In mathematical terms, the classes that were identified
empirically in Ref.~\cite{Bassler04} are group orbits.
We illustrate the mapping for the sixteen $k=2$ functions
in figure 2, where the rotational plus parity
symmetries of the functions belonging to each of the four classes are obvious.
In figure 3 we illustrate one representative cube for each of the 14 function
classes in $k=3$.  The remaining functions in each class are obtained by
applying all possible rotations plus parity to the cube.
In the hypercube representation, 
the canalization properties of a Boolean function 
correspond to the fraction of homogeneous hypersurfaces. That is,
for a Boolean function with $k$ inputs $P_n$ is proportional to the fraction
of the $k-n$ dimensional hypersurfaces that have all vertices the same.
For the two classes with $P_1=1/2$ in figure 2, 2 of the 4 one-dimensional sides are
uniformly coloured.

\begin{figure}
\includegraphics[width=0.5\textwidth]{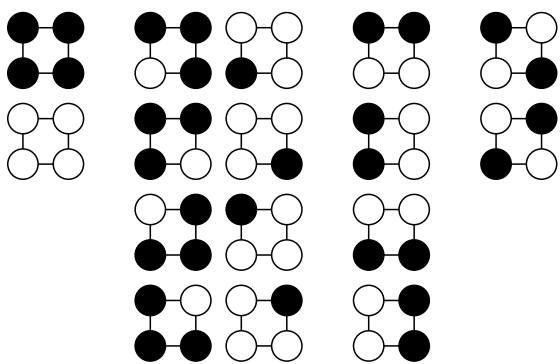}
\caption{Representation of the sixteen $k=2$ functions on Ising squares.
The functions are grouped into four classes.  The members of each class
are clearly related by symmetry operations on the square plus parity.}
\end{figure}

\begin{figure}
\includegraphics[width=0.5\textwidth]{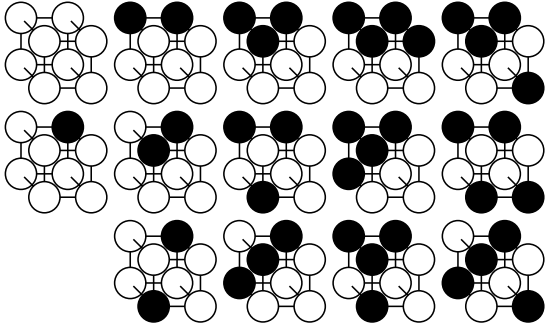}
\caption{A single representative Ising cube mapping of each of the 14 classes 
in $k=3$.}
\end{figure}

We can now employ results from group theory to obtain information about
the class structure of functions 
at all values of $k$, not merely those values of $k$ which are
small enough to permit direct inspection of all functions.
The total number of classes for a given $k$ can be obtained by an application
of 
the orbit-counting theorem,
\begin{equation} 
P_G(x_1,x_2,...)=\frac{1}{|G|}\sum_{g \in G}|X^g|
\end{equation}
Here, the symmetry group $G$ 
of the set $X$ contains $|G|$ symmetry operators $g$, which
together include all transformations of the hypercube onto itself.
The set of elements in $X$ that are left invariant by $g$ is
denoted as $X^g$.
In order to find $X^g$, note that
the mapping of all $k$-hypercube vertices onto themselves by a given
symmetry operator $g$ can be written as a permutation of the vertex
numbers.  As a result, each operator $g$ can be
expressed in terms of its cycle structure $x_1^{b_1}x_2^{b_2}...x_n^{b_n}$,
where $n=k$.  This notation indicates that $g$ contains $b_1$ cycles of
length 1, $b_2$ cycles of length 2, ... $b_n$ cycles of length $n$ 
\cite{King81}.
For example, the $k=2$ permutation (14)(2)(3) has the cycle representation
$x_1^2x_2$ since it has 2 cycles of length 1 and a single cycle of length 2.
To apply 
the orbit-counting theorem, we must first construct all of the operators
of our group, sum the number of functions {\it left invariant} by each
operator (the fixed points of that operator), 
and divide by the total number of operators.

The number of functions in, or size of, a class, is given 
by the number of elements in the group divided by the number of elements in
the isometry
group of the functions in the class. 
The isometry group of a class 
is the subgroup of the full group that
describes the symmetry of a function in the class.  
Note that the particular
isometry group will vary from function to function in the class, 
but the size of the isometry group
will remain invariant.

First, consider the number of
symmetry operations $|G|$ in our group, which is the $k$-hypercube crossed
with parity.
The symmetry group for the $k$-hypercube is isomorphic to the hyperoctahedral
group $O_n$ with $n=k$, 
which has $n!2^n$ symmetry transformations \cite{Edwards00}.
As an example, there are eight operators
for the $k=2$ square.  There is one operator with cycle structure $x_1^4$:
(1)(2)(3)(4); two operators with cycle structure $x_4$:  
(1243) and (3421);  three operators with cycle structure $x_2^2$: 
(12)(34), (13)(24), and (14)(23); and two operators with cycle structure
$x_1^2x_2$: (14)(2)(3) and (23)(1)(4).
When these operators are combined with parity,
which doubles the number of symmetry operators, we obtain a
total of $|G|=16$ operators.
For each operator without parity, the number of functions left invariant 
is
equal to $2^{N_c}$, where $N_c=\sum_{i=1}^{k}b_i$ is the total 
number of cycles in the operator.  
Parity must be treated
separately; no functions are left invariant by
the parity operator with any 
$k$-hypercube operator
containing at least one cycle of length 1.  
Thus there are $2^{N_p}$ functions left invariant 
for the eight operators which include parity, 
where
$N_p=(1-\Theta(b_1))\sum_{i=1}^k b_i$ and  $\Theta$ is the
Heaviside step function.
Applying 
the orbit-counting theorem 
produces the correct
number of classes, but {\it only} if parity is included.  In the case of
$k=2$ without parity,
$P_G=(1/8)(2^4+2(2)+3(2^2)+2(2^3))=6$ classes.
Including parity gives
$P_G=(1/16)(2^4+2(2)+3(2^2)+2(2^3)+2(2)+3(2^2))=4$ classes, which is the
correct result.

We now face the task of identifying all operators $g$ for the 
$k$-hypercube.  This can be performed by simple inspection in $k=2$ and $k=3$,
but it becomes more complicated to keep track of higher-dimensional rotation
symmetries.  Fortunately, this problem was solved in the middle of the
last century, when Harrison \cite{Harrison} derived a formula that 
produces the complete cycle representation for all $k$ in the group of 
interest to us, called the ``Zyklenzeiger'' in the notation of 
Ref.~\cite{Harrison}.  We have used this formula to obtain the cycle
representations through $k=5$, shown in table 3.  The number of classes
for each $k$, $P_G$, is also listed in table 3.  
Clearly, obtaining the properties of 
the classes by inspection is not feasible by the time $k=5$.

\begin{table}
\caption{Cycle polynomials for $k=1$ through 5 and the number of
classes $P_G$ for each $k$.}
\begin{indented}
\lineup
\item[]\begin{tabular}{@{}clc}
\br
k & Cycle polynomial & \0\0$P_G$ \\ 
\mr
1 & $(1/2)(x_1^2+x_2)$ & \0\0\0\0\02 \\ 
2 & $(1/8)(x_1^4+3x_2^2+2x_1^2x_2+2x_4)$ & \0\0\0\0\04 \\ 
3 & $(1/48)(x_1^8+13x_2^4+8x_1^2x_3^2+8x_2x_6+6x_1^4x_2^2+12x_4^2)$ & \0\0\0\014
\\ 
4 & $(1/384)(x_1^{16}+12x_1^8x_2^4+51x_2^8+12x_1^4x_2^6+32x_1^4x_3^4+48x_1^2x_2x_4^3+84x_4^4$ & \\
& $+96x_2^2x_6^2+48x_8^2)$ & \0\0\0238 \\ 
5 & $(1/3840)(x_1^{32}+384x_{10}^3x_2+20x_1^{16}x_2^{8}+60x_1^8x_2^{12}+231x_2^{16}+80x_1^8x_3^8$ & \\
& $+320x_{12}^2x_4^2
+240x_1^4x_2^2x_4^6+240x_2^4x_4^6+520x_4^8+384x_1^2x_5^6$ & \\
& $+160x_1^4x_2^2x_3^4x_6^2+720x_2^4x_6^4+480x_8^4)$ & 698635 \\
\br
\end{tabular}
\end{indented}
\end{table}

What we are interested in is not simply how many different classes are
present,  but also the size of each class, and 
the structure, particularly the canalization properties,
of the functions belonging to them.
For example, how many classes are there
which have the same internal inhomogeneity $p$?  To find this, we use
an application of P\' olya's theorem \cite{Polya} 
which is frequently used in isomer chemistry \cite{Balasubramanian82}.
In isomer chemistry, for a molecule composed of exactly two different
types of atoms, the terms $A$ and $B$ can be used to represent the different 
types of atoms.  In our case, $A$ and $B$
represent 0 and 1, such that either $A=0$ and $B=1$, or $B=0$ and $A=1$.
Using the generating polynomial, substitute in a term
of the form $A^a+B^a$ for each $x_a$.  Divide the result by the total number
of operators, including parity.  Then, drop all terms in the result where
the exponent of B exceeds that of A, as these terms are already accounted
for by parity.  The multiplicity of each term indicates how many of the classes
are of that form. For example, representatives of classes of the 
form $A^2B^2$ are 1010 and 1100.
This gives us the desired result
of how many classes $N_h$ 
there are for each value of the internal homogeneity $p$.
Since we also know that the total number $N_f(m,n)$ of functions of the form
$A^mB^n$ is simply $N_f(m,n)=(2-\delta_{m,n})(2^k)!/(m!n!)$, where $m+n=2^k$,
we can estimate the number of functions
in each class by the average size of a class, 
$\langle S_c\rangle= N_f/N_h$.  
The class structure and
average class size for $k=1$ through 5 are listed in tables 4 through 7.
We note that the actual size $S_c$ of each class is given by the number of
operators that preserve the symmetry of that particular function class.
Thus, as discussed earlier, the maximum class size $S_c^{max}$ 
is equal to the total number of operators 
\begin{equation}
S_c^{max}=k!\;2^{k+1} . 
\end{equation}
$S_c^{max}=16$ for $k=2$, 96 for $k=3$, 768 for $k=4$, and 7680 for $k=5$.
This is consistent with the average class sizes which we obtain.

We also note that isomer chemistry provides a simple means for determining
whether two randomly selected functions belong to the same class.  Construct
the adjacency matrix for the $k$-hypercube.  Along the diagonal, place the
values $A$ or $B$ corresponding to the colours of the
vertices of one of the functions
under consideration, and then find the determinant of the resulting matrix.
Each function class has a unique determinant, so performing this
procedure on both functions provides an immediate
test of whether the two functions fall into the same class.

\begin{table}
\caption{Class structure for $k=2$.}
\begin{indented}
\lineup
\item[]\begin{tabular}{@{}cll}
\br
Class type & $N_h$ & $\langle S_c\rangle$ \\ 
\mr
$A^4$ & 1 & 2\\ 
$A^3B$ & 1 & 8\\
$A^2B^2$ & 2 & 3\\
\br
\end{tabular}
\end{indented}
\end{table}

\begin{table}
\caption{Class structure for $k=3$.}
\begin{indented}
\lineup
\item[]\begin{tabular}{@{}cll}
\br
Class type & $N_h$ & $\0\langle S_c\rangle$ \\ 
\mr
$A^8$ & 1 & \02\\ 
$A^7B$ & 1 & 16\\
$A^6B^2$ & 3 & 18.667\\
$A^5B^3$ & 3 & 37.333\\
$A^4B^4$ & 6 & 11.667\\
\br
\end{tabular}
\end{indented}
\end{table}

\begin{table}
\caption{Class structure for $k=4$.}
\begin{indented}
\lineup
\item[]\begin{tabular}{@{}cll}
\br
Class type & $N_h$ & \0$\langle S_c\rangle$ \\ 
\mr
$A^{16}$ & \01 & \0\02\\ 
$A^{15}B$ & \01 & \016\\
$A^{14}B^2$ & \04 & \060\\
$A^{13}B^3$ & \06 & 186.667\\
$A^{12}B^4$ & 19 & 191.58\\
$A^{11}B^5$ & 27 & 323.56\\
$A^{10}B^6$ & 50 & 320.32\\
$A^9B^7$ & 56 & 408.57\\
$A^8B^8$ & 74 & 173.9\\
\br
\end{tabular}
\end{indented}
\end{table}

\begin{table}
\caption{Class structure for $k=5$.}
\begin{indented}
\lineup
\item[]\begin{tabular}{@{}cll}
\br
Class type & \0\0$N_h$ & \0\0$\langle S_c\rangle$ \\ 
\mr
$A^{32}$ & \0\0\0\0\01 & \0\0\02\\ 
$A^{31}B$ & \0\0\0\0\01 & \0\064\\
$A^{30}B^2$ & \0\0\0\0\05 & \0198.4\\
$A^{29}B^3$ & \0\0\0\010 & \0992\\
$A^{28}B^4$ & \0\0\0\047 & 1530.2\\
$A^{27}B^5$ & \0\0\0131 & 3074.4\\
$A^{26}B^6$ & \0\0\0472 & 3839.8\\
$A^{25}B^7$ & \0\01326 & 5076.7\\
$A^{24}B^8$ & \0\03779 & 5566.7\\
$A^{23}B^9$ & \0\09013 & 6224.1\\
$A^{22}B^{10}$ & \019963 & 6463.2\\
$A^{21}B^{11}$ & \038073 & 6777.7\\
$A^{20}B^{12}$ & \065664 & 6877.2\\
$A^{19}B^{13}$ & \098804 & 7031.6\\
$A^{18}B^{14}$ & 133576 & 7058.7\\
$A^{17}B^{15}$ & 158658 & 7131.3\\
$A^{16}B^{16}$ & 169112 & 3554.3\\
\br
\end{tabular}
\end{indented}
\end{table}

To show that canalization remains important even as $k$ increases,
we measured the average number of homogeneous $d$-dimensional sides
present in a series of randomly generated functions for different
$k$.
We denote the number of $d$-dimensional homogeneous sides 
(which produce canalization) that a 
$k$-dimensional Boolean function has as $C(d,k)$.  
The total number of
$d$-dimensional sides is 
\begin{equation}
N_d(k)=\frac{2^{k-d}k!}{(k-d)!d!}.
\end{equation}
Note that the canalization properties $P_n$ discussed in the
beginning of this paper are related to $C(d,k)$ 
as $P_n=C(k-n,k)/N_{k-n}(k)$.
In figure 4 we plot the average fraction of homogeneous $d$ dimensional
sides, $c_d=\langle C(d,k)\rangle/N_d(k)$, for 
$d=1$ through $4$ and $k=2$ through 8, obtained
numerically.  We sampled up to $1\times 10^8$
functions generated with $p=0.5$.  For the case of $d=1$, shown in 
figure 4(a), on average over 50\% of the sides of the hypercube
are uniformly coloured even for $k=8$, indicating a significant amount of
partial canalization.  As $d$ increases, $c_d$ drops considerably, as
illustrated in figure 4(b-d) for $d=2$, 3, and 4.  Thus, the most
prevalent type of partial canalization is that associated with
homogeneous $d=1$ sides.

\begin{figure}
\includegraphics[width=0.7\textwidth]{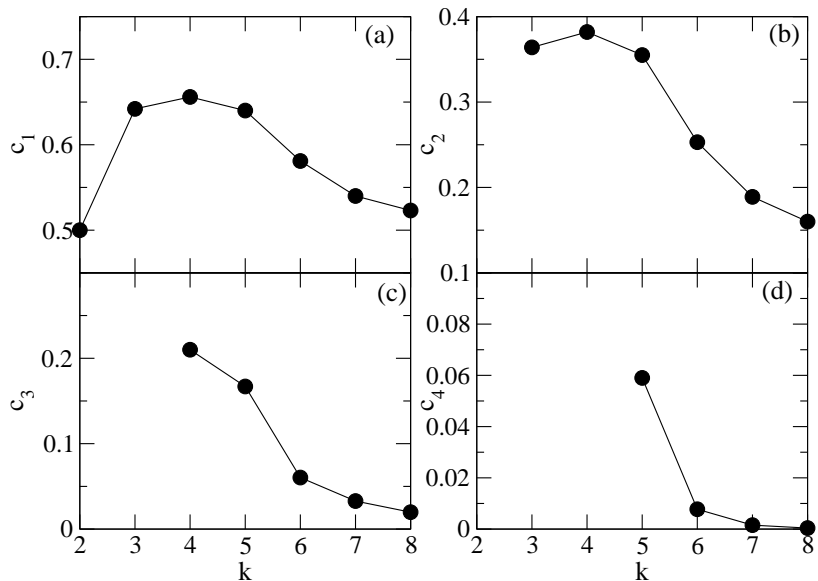}
\caption{
Average fraction $c_d$ of homogeneous $d$-dimensional sides 
in randomly selected Boolean functions versus
$k$ for (a) $d=1$, (b) $d=2$, (c) $d=3$, and (d) $d=4$.
}
\end{figure}


The mapping of Boolean functions onto $k$-hypercubes 
we have described here provides a means of
constructing $k+1$ functions recursively from pairs of $k$ functions.
A $k+1$ function can be composed by stacking together two $k$ functions.
Depending on the symmetry properties of the two $k$ functions chosen,
there may be only one possible class of $k+1$ functions that can be
constructed from those $k$ functions, or there may be several classes that 
depend on the relative orientation of the $k$ functions when they are
stacked together.  This allows us to bound the amount of
canalization present.
When we assemble a $k+1$ function out of two $k$ functions, we must
have
\begin{equation}
N_d(k+1)\ge C(d,k+1) \ge \sum_{i=1}^2 C_i(d,k).
\end{equation}
The lower bound is obtained from the fact that
the value of $C(d,k+1)$ must be at least as large as the 
sum of the values $C_i(d,k)$ of the two functions that have been combined.  
This is simply a consequence of the fact that homogeneous 
$d$-dimensional sides 
cannot be destroyed as a result of combining two functions.  The new sides that
are added when the functions are joined may or may not be homogeneous,
depending on which two $k$ functions are combined and 
how they are oriented with
respect to each other.  It is possible that none of the new sides would
be homogeneous, in which case the lower limit of Eq. (4) would apply.
The internal homogeneity $p(k+1)$ of the composite function is given
simply by
\begin{equation}
p(k+1)=\frac{1}{2}\sum_{i=1}^2 p(k) .
\end{equation}

\section{Discussion}

Now that we have used the mapping of the functions onto $k$-hypercubes
in order to obtain information about the class structures of the
functions for several values of $k$, we can make some observations
regarding how prevalent partial canalization is among all possible
functions.  Previous estimates of the fraction of canalizing functions
indicated that canalization was of less and less importance as $k$
increased.  These estimates used a very narrow definition of canalization,
however.  Rather than counting the number of partially canalizing functions,
consider the number of completely uncanalizing functions.  These functions
have the property that they are sensitive to all values of all inputs.
There are exactly 2 such functions for each $k$, regardless of $k$.
{\it All} of the remaining functions are at least partially canalizing.
This means that partial canalization completely dominates the classes of
functions, especially as $k$ increases.

The rampant occurrence of partial canalization has important implications
for recent work on mapping of genetic regulatory networks.  The experiments
typically 
map only those connections between genes which are active in the native
state of the organism.  Here, ``active'' means that a change in one gene
directly affects the second gene.  This technique will {\it not detect}
many of the partially canalizing interactions 
that could exist between genes.  In the 
case where the partial canalization is of the form that a gene completely
ignores one or more of its inputs, 
the actual value of $k$ for that gene is larger than
the apparent value of $k$.  This could potentially impact the distributions
of $k$ that have been extracted from experimental measurements.
A far more dangerous case is a partially canalizing interaction between
genes in which a gene ignores one or more inputs when a canalizing input
has a value of 1 (for example), but responds to the other inputs when that
same canalizing input has a value of 0.  If the gene ignores its inputs in
the native state, the connection between that gene and its ignored inputs
will not be detected experimentally.  Suppose that the canalizing gene is
identified as causing a disease state.  Consultation of the experimentally
determined genetic network map indicates that this gene does not appear to
control anything else of importance.  If, however, the canalizing gene
is treated and switches to the state opposite from its canalizing value,
the gene that received the canalizing input will suddenly start to respond
to the values of its other inputs.  This could result in unexpected
side effects or worse effects.  Thus, from a purely combinatorial point of
view, it is important to consider all possible interactions between genes,
and not merely those which are expressed in the native state.

The natural predominance of canalization as $k$ increases suggests that the 
canalization observed experimentally could be due simply to the high fraction
of the available Boolean functions which are canalizing, rather than
evolutionary pressure to develop canalizing
functions. It is, however, 
unclear how much canalization is present in real genetic
regulatory networks, as discussed in Ref.~\cite{Grefensteffe}. 
It is possible that there is in fact
a special evolutionary preference
for canalization, which could result in real networks 
having even higher levels of canalization than would be expected from 
random selection at increasing $k$.
In order to answer this question it would be necessary to
measure the {\em excess
canalization}, which is the difference between 
the $P_n$s observed in
real networks and that in random networks \cite{Liu05}.
The existing experimental data on 
genetic regulatory networks is not extensive enough to determine whether
an interaction between genes is 
canalizing or partially canalizing.
As noted above, the difference between the two types of interactions can 
become important when the network is perturbed away from its native state.  
More experimental
work is needed in order to determine the prevalence of canalization
and/or partial canalization in actual genetic regulatory
networks.  The Boolean models can
offer guidance in determining how likely it would be to observe
any type of canalization in a random network.

\section{Conclusion}

In conclusion, we have used a mapping of the Boolean functions in the Kauffman
model for genetic regulatory networks onto a $k-$hypercube to obtain
information about the classes into which the functions can be divided.
These classes arise due to geometrical constraints, and can be constructed
by applying all possible rotations of the $k-$hypercube plus parity to
each function.  The classes can be counted and their properties determined
using results from group theory and isomer chemistry.  
We emphasize that
partially canalizing functions completely dominate all possible functions,
particularly for higher $k$.  This indicates that partial canalization should
be extremely common, even in a randomly chosen network, and has implications
for how much information can be obtained in experiments on native state
genetic regulatory networks.

\ack
We thank Min Liu for discussions.
This work was supported by the U.S. DoE under Contract No. W-7405-ENG-36 
(CJOR), the LANL Laboratory Directed Research 
and Development program (CJOR),
and
the NSF through grant No.\ DMR-0427538 (KEB).

\end{document}